\documentstyle[preprint,aps]{revtex}

\begin{document}
\bibliographystyle{prsty}
\draft

\title{Preparation of Schr\"odinger cat states with cold ions beyond the Lamb-Dicke limit}
\author{Mang Feng
\thanks{Electronic address: feng@mpipks-dresden.mpg.de}}
\address{$^{1}$Max-Planck Institute for the Physics of Complex Systems,\\
N$\ddot{o}$thnitzer Street 38, 01187 Dresden, Germany\\
$^{2}$Laboratory of Magnetic Resonance and Atomic and Molecular Physics,\\ 
Wuhan Institute of Physics and Mathematics, Academia Sinica,\\
Wuhan 430071 People's Republic of China}

\date{\today}
\maketitle

\begin{abstract}

A scheme for preparing Schr\"odinger cat (SC) states is proposed beyond the Lamb-Dicke limit in a 
Raman-$\Lambda$-type configuration. It is shown that  SC states can be obtained more efficiently
with our scheme than with the former ones.

\end{abstract}
\vskip 1cm
\pacs{PACS numbers: 42.50.Vk,03.65.-w,32.80.Pi}

\narrowtext

In recent years the system of cold ions moving in a harmonic trap is considered to be a prospective physical setting for
the preparation and investigation of nonclassical states $^{[1-3]}$. It has been proven that, in the limit where coherent 
interaction can dominate over dissipative process,  the model of a cold ion strongly coupled to a harmonic oscillator is 
formally similar to the cavity quantum electrodynamics (QED). 
 
Almost all schemes for the preparation of nonclassical motional states of a trapped ion are based on the Lamb-Dicke 
limit(LDL) under the weak excitation regime(WER), which correspond to the actual case in the present ion-trap 
experiments $^{[4-6]}$. 
The LDL means that the spatial dimensions of the ground motional state are much smaller than the effective wavelength of the 
laser wave, and the WER is the condition that the Rabi frequency describing 
the laser-ion interaction is much smaller than the trap frequency. In the case of the LDL and WER, 
Jaynes-Cummings model(JCM)$^{[7,8]}$ can be used to describe the trapped ion system in the supposition
that the ion is of two levels, the trap's potential can be quantized to be a harmonic oscillator, and
the radiating lasers  can be taken as the classical forms of  standing or traveling waves.
Some techniques developed in the framework of cavity QED based on JCM can be immediately transcribed 
to the ion trap system by taking advantage of the analogy between the cavity QED and the ion trap problem.
Recently, a scheme $^{[2,9]}$ for the preparation of Schr\"odinger cat (SC) states was proposed under the 
strong excitation regime (SER), opposite to the WER. As in the SER the Rabi frequency is very large, the
operation time for the state preparation can be much reduced, which is advantageous to avoid the decoherence. 
On the other hand, we also noted that the case beyond the LDL has been discussed intensively for the rapid 
laser-cooling of the ion $^{[10-12]}$. It has been shown that the laser-ion interaction beyond the LDL 
in the manipulation of the cold ions is helpful for reducing  the noise in the trap, loosing the confinement 
of the trap and improving the cooling rate$^{[12]}$. However, as far as we know,
no specific  proposal has been put forward so far for preparing nonclassical states of the ion beyond the LDL. 

In this letter, we try to prepare the SC states $^{[13]}$ under the WER, but beyond the LDL. 
The SC state, i.e., the superposition of macroscopically distinguishable states, has been drawn much attention
over past several decades due to both the SC paradox, i.e., a Schr\"odinger's thought experiment, and the possibility 
of experimental realization in the mesoscopic system. Under both the LDL and WER, there have been some proposals $^{[14-16]}$ 
for the preparation of  SC states with trapped cold ions. Experimentally, SC states have been obtained in NIST
group with single cold $^{9}Be^{+}$$^{[5]}$. We will show that, beyond the LDL, the preparation of SC states can be 
made more rapidly and simply than those within the LDL. From the viewpoint of decoherence, it may be of importance for 
the experimental implementation and measurement due to the reduction of the operation time.
 
We investigate the situation that the single ultracold ion radiated by lasers in the Raman-$\Lambda$-type
configuration$^{[4]}$. The electronic structure is employed with two lower levels $|e>$ and $|g>$ coupled 
to a common upper state $|r>$, and the two lasers with frequencies $\omega_{1}$ and 
$\omega_{2}$ respectively are assumed to propagate along opposite direction. For a 
sufficiently large detuning to the level $|r>$, $|r>$ may be adiabatically 
eliminated, and what we have to treat is an effective two-level system, in which the 
lasers drive the electric-dipole forbidden transition $|g>$$\leftrightarrow |e>$.
The dimensionless Hamiltonian of such a system in the frame rotating with the 
effective laser frequency $\omega_{l}$$(=\omega_{1}-\omega_{2})$ can be written as$^{[9]}$
\begin{equation}
H=\frac {\Delta}{2} \sigma_{z} + a^{+}a
+\frac {\Omega}{2}[\sigma_{+}e^{i\eta (a^{+}+a)}+\sigma_{-}e^{-i\eta (a^{+}+a)}]
\end{equation}
where the detuning $\Delta=(\omega_{0}-\omega_{l})/\nu$ with $\omega_{0}$ 
being the transition frequency of two 
levels of the ion, and $\nu$ the frequency of the trap. $\Omega$ is the Rabi 
frequency and $\eta$  the effective Lamb-Dicke parameter given by $\eta=\eta_{1}+\eta_{2}$ with 
subscripts denoting the counterpropagating laser field. $\sigma_{i}$ ($i=\pm, z$) are Pauli operators, and
$a^{+}$ and $a$ are operators of creation and annihilation of the phonon 
field, respectively. The notations '+' and '-' in front of $i\eta (a^{+}+a)$ 
correspond to the absorption of a photon from one beam followed by emission into the other beam and vice versa,
respectively. $\nu$ is generally supposed to be much greater 
than the atomic decay rate, called the strong confinement limit, for neglecting the effect 
of the atomic decay. 
We first perform following unitary transformations on Eq.(1), that is$^{[17]}$
\begin{equation}
H^{I}=THT^{+}=\frac {\Omega}{2}\sigma_{z} + a^{+}a - i\xi(a^{+}-a)\sigma_{x} -\epsilon \sigma_{x} + \xi^{2}
\end{equation}
where $T=\frac {1}{\sqrt{2}}\pmatrix{D^{+} & D\cr -D^{+}& D}$ 
with $D=e^{i\xi(a^{+}+a)}$, $\xi=\eta/2$, $\epsilon=\Delta/2$,
and $\sigma_{x}=\sigma_{+}+\sigma_{-}$. As we suppose $\Omega \ll 1 \le \xi$,  Eq.(2)
can be reduced to 
\begin{equation}
H^{I}= a^{+}a - i\xi(a^{+}-a)\sigma_{x}-\epsilon \sigma_{x} + \xi^{2}
\end{equation} 
where the detuning term $\epsilon$ is retained due to  no special requirement on it. 
Therefore, the time evolution operator in the original representation is
$$\hat{U}(t)=T^{+}\exp(-iH^{I}t)T$$
\begin{equation}
=\frac {1}{2} e^{-i\xi^{2}t}\pmatrix{D & -D\cr D^{+}& D^{+}}
e^{-it[a^{+}a - i\xi(a^{+}-a)\sigma_{x}-\epsilon \sigma_{x}]}\pmatrix{D^{+} & D\cr -D^{+}& D}.
\end{equation} 
Direct algebra on Eq.(4) yields
$$\hat{U}(t)=\frac {1}{2} e^{-i\xi^{2}t}\pmatrix{e^{-ia^{+}at}De^{-\xi t(a^{+}-a)} &
 -e^{-ia^{+}at}De^{-\xi t(a^{+}-a)} \cr e^{-ia^{+}at}D^{+}e^{\xi t(a^{+}-a)} 
 & e^{-ia^{+}at}D^{+}e^{\xi t(a^{+}-a)}}\times$$ 
$$ \pmatrix{cosh [\xi (a^{+}-a)t] & 
 -sinh [\xi (a^{+}-a)t] \cr -sinh [\xi (a^{+}-a)t] & cosh [\xi (a^{+}-a)t]}\times$$
\begin{equation}
\pmatrix{\cos [\frac {\xi}{2}t^{2}(a^{+}+a)] & 
 -i\sin [\frac {\xi}{2} t^{2}(a^{+}+a)] \cr -i\sin [\frac {\xi}{2} t^{2}(a^{+}+a)] & \cos [\frac {\xi}{2} t^{2}(a^{+}+a)]}
\pmatrix{\cos\epsilon t & i\sin\epsilon t \cr i\sin\epsilon t & \cos\epsilon t}
\pmatrix{D^{+} & D\cr -D^{+}& D}
\end{equation} 
in which we have used Baker-Campbell-Hausdorff theorem, and formulas $e^{iA\sigma_{x}}=\cos A + i\sigma_{x}\sin A$ and 
$e^{A\sigma_{x}}=cosh A + \sigma_{x}sinh A$.

Consider that the ion has been laser-cooled to the dark state $|g>|0>$ beyond the LDL$^{[10-12]}$, where we denote 
the electronic and motional 
ground states by $|g>$ and $|0>$ respectively. Defining $|e>=\pmatrix{1\cr 0}$ and $|g>=\pmatrix{0\cr 1}$,  a laser pulse 
$\hat{V}=\frac {1}{\sqrt{2}}\pmatrix{1& 1 \cr -1 & 1}$ applied on the ion will yield the 
state $\Psi_{1}=\frac {1}{\sqrt{2}}(|e>+ |g>)|0>$. Then performing $\hat{U}(t)$ on $\Psi_{1}$, we have
the superposition of coherent states correlated with the internal states of the ion
\begin{equation}
\Psi_{2}=\frac {1}{\sqrt{2}}e^{-i\xi^{2} t}[e^{-i\epsilon t}|e>|i\frac {\xi}{2} t^{2}e^{-it}>
+ e^{i\epsilon t}|g>|-i\frac {\xi}{2} t^{2}e^{-it}>].
\end{equation} 
Finally, we apply $\hat{V}$ once again, which produces 
\begin{equation}
\Psi_{3}=\frac {1}{\sqrt{2}}e^{-i\xi^{2} t}(\Phi_{+}|e> + \Phi_{-}|g>)
\end{equation}
with the SC states $\Phi_{\pm}=\frac {1}{\sqrt{2}}(e^{i\epsilon t}|-i\frac {\xi}{2} 
t^{2}e^{-it}>\pm e^{-i\epsilon t|}|i\frac {\xi}{2} t^{2}e^{-it}>)$.

Let us take more specific consideration on Eq.(7). To measure the SC states perfectly, we can use
the technique of electronic shelving amplification$^{[18]}$. By introducing the fourth electronic level $|f>$ of the ion, and a weak laser
beam resonant with the transition of $|g>\rightarrow |f>$, we can obtain
$\Phi_{+}$ perfectly corresponding to no 
fluorescence in observation. However, to obtain a perfect $\Phi_{-}$, we have to modify the 
last step of above preparation process, that is, replacing $\hat{V}$ with $\hat{V}'=\frac {1}{\sqrt{2}}\pmatrix{1& -1 \cr 1 & 1}$. 
Then we have 
\begin{equation}
\Psi^{'}_{3}=\frac {1}{\sqrt{2}}e^{-i\xi^{2} t}(-\Phi_{-}|e> + \Phi_{+}|g>).
\end{equation} 
So the perfect $\Phi_{-}$ can be obtained similarly in the absence of fluorescence. 
Another problem is that, $\Phi_{\pm}$ should be macroscopic in the sense that the component states 
$|-i\frac {\xi}{2} t^{2}e^{-it}>$ and $|i\frac {\xi}{2} t^{2}e^{-it}>$ are distinguishable. Unfortunately, if 
$t=k\pi$ with $k=0,1,2,\cdots$, SC states
$\Phi_{\pm}=\frac {1}{\sqrt{2}}[e^{i\epsilon k\pi}|-(-1)^{k}i\frac {\xi}{2} k^{2}\pi^{2}>\pm e^{-i\epsilon k\pi}
|(-1)^{k}i\frac {\xi}{2} k^{2}\pi^{2}>]$
 can not be observed directly since its probability distribution in the position
representation has only one peak centred at $<R>=0$$^{[2]}$. So we had better choose the time 
$t=\frac {1}{2}(2k+1)\pi$ with $k=0,1,2,\cdots$. The SC state at this moment is of the form 
\begin{equation}
\Phi_{\pm}=\frac {1}{\sqrt{2}}[e^{i \frac {\epsilon}{2}(2k+1)\pi}|-(-1)^{k}\frac {1}{8}\xi (2k+1)^{2}\pi^{2}>\pm 
e^{-i \frac {\epsilon}{2}(2k+1)\pi}|(-1)^{k}\frac {1}{8}\xi (2k+1)^{2}\pi^{2}>]
\end{equation}
which has two maxima in the position representation centred at 
$<R>\sim \pm\frac {1}{8}\xi (2k+1)^{2}\pi^{2}$.  As distinguishing the two component coherent states needs the two peaks to be spatially
separated by more than a wavelength, we should wait for at least $t\ge \sqrt{2\pi/\xi}$  when applying $\hat{U}(t)$ on the ion.
 
The form of Eq.(7) is very similar to the experimental results in Ref.[5]. However, as the laser-cooling of the ion beyond the LDL 
to the motional ground state is more rapid than the case under the LDL$^{[12]}$, and the procedure of preparing the SC state in our scheme 
is simpler than that in Ref.[5], our scheme is obviously more efficient.    
In fact, comparing with former various schemes, our scheme is also more efficient in the preparation of the SC state. In Refs.[2,9],
the laser pulses should be applied on the ion alternatively from opposite directions for many times in order to obtain 
a observable SC state. In
contrast, we only have three steps of the laser-ion interaction, i.e.,  $\hat{V}\hat{U}(t)\hat{V}$ or $\hat{V}\hat{U}(t)\hat{V}'$, and 
a suitable choice of $t$. In
Ref.[14], the SC state is obtained as the final result of the spontaneous emission process. So the state preparation will take a long time if 
the trapped ion is nearly isolated from the external environment. Moreover, as Refs.[15,16] describe the process under the LDL and WER,
 it is easily found that  preparing observable SC states with those schemes is also time-consuming.    
On the contrary, in our scheme, as $\eta$ is large, and the values of the component coherent states are proportional to $t^{2}$, 
the time for preparing a observable SC state is much shorter than that in above schemes.  

More specifically, even if we neglect the speed-up in the laser-cooling of the trapped ion beyond the LDL, following
simple numerical estimates for the preparation time of the observable SC states can also show the advantage of our scheme. With 
current experimental parameters of 
$\eta=0.202$, (dimensionless) $\Omega=0.1$ and $\nu=10^{7} Hz$$^{[4,5]}$, we obtain that the dimensionless time 
for preparing a observable SC state with the scheme in [2] is $\frac {\pi}{4}(\sqrt{\frac {\pi}{2\eta^{2}}}-1)=4.09$ 
in the supposition that  the delay time between any two operations of laser pulses can be omitted, and the time corresponding to 
[16] is $\frac {\pi e^{\eta^{2}/2}}{\eta\Omega}=159$.
If we choose the scheme in [14] and the cold ion $^{9}Be^{+}$,  the time will be much longer since 
the metastable level of $^{9}Be^{+}$ is $1\sim 10$ seconds. So the minimum dimentionless time is $10^{7}$. 
In contrast, with our scheme, $t \sim\sqrt{4\pi/\eta}=2.51$ if $\eta=2.0$. In fact, 
$\eta$ can be maximized to 3.0 $^{[10,12]}$, so $t$ can be minimized to 2.05. 

In summary, a simple but efficient scheme for preparing the SC states of motion of a cold trapped ion  has been proposed 
based on the ion under the WER but beyond the LDL.
 As far as we know, it is the first proposal for the preparation of
nonclassical motional  states of the cold ions  beyond the LDL.  
As the trapped ions beyond the LDL are less tightly confined and more rapidly laser-cooled, which meets the 
requirement of the ion trap quantum computing$^{[19]}$,
the investigation on the laser-ion interaction beyond the LDL is of importance. We also note a proposal using SC states
to be the robust qubits for the quantum computing$^{[20]}$. Therefore we believe the present work would be helpful 
for the future exploration of the ion-trap system beyond the LDL, although the experimental work in this 
respect has not been reported yet.

The work is partly supported by the Chinese National Natural Science
Foundation under Grant No.19904013.

\end{document}